\documentclass[numreferences]{kluwer}

\usepackage{mathbbol, klucite, bm}
\renewcommand{\v}[1]{\mathbf{#1}}

\renewcommand{\d}[0]{\partial}

\newcommand{\bl}[0]{\bullet}

\newcommand{\ket}[0]{\rangle}

\newcommand{\bra}[0]{\langle}

\newcommand{\na}[0]{\nabla}

\newcommand{\RR}[0]{\mathbb{R}}

\newcommand{\tr}[1]{{\rm tr}\nolinebreak\hspace{2pt}\nolinebreak #1}

\newlength{\minus}
\newcommand{\ms}[0]{\hspace{\minus}}

\newenvironment{mat}[1]{\begin{array}{@{}*{#1}{r@{}l}@{}}}{\end{array}}
\newtheorem{theorem}[subsection]{Theorem}

\begin{document}
\begin{article}
\begin{opening}
\title{Application of Sharafutdinov's Ray Transform in Integrated
Photoelasticity}
\author{H. \surname{Hammer} \email{H.Hammer@umist.ac.uk}}
\author{B. \surname{Lionheart} \email{Bill.Lionheart@umist.ac.uk}}
\institute{Department of Mathematics, \\ UMIST, PO Box 88, \\
 Manchester M60 1QD, UK}
\runningtitle{Ray Transform in Integrated Photoelasticity}
\runningauthor{Hammer e.a.}
\begin{ao}
Kluwer Prepress Department\\
P.O. Box 990\\
3300 AZ Dordrecht\\
The Netherlands
\end{ao} 
\begin{abstract} 
We explain the main concepts centered around Sharafutdinov's ray
transform, its kernel, and the extent to which it can be inverted. It
is shown how the ray transform emerges naturally in any attempt to
reconstruct optical and stress tensors within a photoelastic medium
from measurements on the state of polarization of light beams passing
through the strained medium. The problem of reconstruction of stress
tensors is crucially related to the fact that the ray transform has a
nontrivial kernel; the latter is described by a theorem for which we
provide a new proof which is simpler and shorter as in Sharafutdinov's
original work, as we limit our scope to tensors which are relevant to
Photoelasticity. We explain how the kernel of the ray transform is
related to the decomposition of tensor fields into longitudinal and
transverse components. The merits of the ray transform as a tool for
tensor reconstruction are studied by walking through an explicit
example of reconstructing the $\sigma_{33}$-component of the stress
tensor in a cylindrical photoelastic specimen. In order to make the
paper self-contained we provide a derivation of the basic equations of
Integrated Photoelasticity which describe how the presence of stress
within a photoelastic medium influences the passage of polarized light
through the material.
\end{abstract}

\classification{AMS Mathematics Subject Classifications
(2000)}{53C65, 53C80, 44-02, 44A12}

\end{opening}
\section{Introduction}
A transparent optical medium which initially is isotropic and
homogeneous with respect to light propagation may become birefringent
when subjected to external strain, an effect which is known as {\it
Photoelasticity} \cite{Frocht,CokerFilon}. The spatially varying
tensor of refraction then reflects the presence of stresses within the
material, opening up the possibility of examining internal stresses by
means of the change in the state of polarization of light beams
passing through the strained medium. The reconstruction of local
optical and stress tensors from the set of data collected for all
possible directions, and locations, of light beams passing through the
specimen is called {\it Integrated Photoelasticity} \cite{Aben1966a}.

Numerous efforts have been made in the past to tackle the problem of
Integrated Photoelasticity
\cite{Aben1966a,Aben1979,TheocarisEA,AbenGuillemet,AbenEA1989a,AndrienkoEA1992a,%
AndrienkoEA1992b,AndrienkoEA1994a,kubo1997a,Schupp1999a}. While the
reconstruction of the optical tensors in the two-dimensional problem
is well-established (for an overview see \cite{TheocarisEA}), the
three-dimensional (3D) case so far has been solved only for special
cases involving {\it a priori} assumptions about the symmetry of the
3D stress distribution, the rotation of the principal axes of the
stress tensor, or the strength of the anisotropy of the dielectric
tensor. The most general case of 3D tensor tomography with arbitrarily
shaped bulk matter and no restriction on the degree of anisotropy is
unsolved. The source of difficulty is the fact that the Radon
transform which provides a well-established formalism for the
reconstruction of scalar fields can no longer be utilized if tensor
fields are involved.

In \cite{Sharafutdinov}, Sharafutdinov has proposed a generalization
of the Radon transform to the case of symmetric tensor fields of
arbitrary degree $m$ defined on a, possibly curved, Riemannian
manifold of arbitrary dimension $n$; he has termed his construction
the {\it ray transform}. Amongst many other aspects, his work
encompasses the examination of the formal structure of the ray
transform, its relation to the Fourier transform, its non-trivial
kernel and the problem of inversion of the ray transform given the
fact that the kernel is non-zero. Sharafutdinov's contribution
undoubtedly opens up the right path to the goal of solving the general
problem of Integrated Photoelasticity; however, since his work aims to
tackle the most general case of symmetric tensors of arbitrary degree
$m$ in $n$ dimensions, his formalism necessarily exhibits a degree of
complexity which may be undesirable for practitioners who wish to put
his theory into a specific physical or engineering context. It is
therefore an important task to lay out his results in a way which
focuses on the special case of symmetric ($m=2$) tensor fields, and
vector ($m=1$) fields, defined in a three-dimensional Euclidean space
which is to be identified with the bulk material of the medium, and to
study the merits of the ray transform as a tool for tensor
reconstruction within the specific framework of Photoelasticity.

In the present paper we have attempted to provide some steps in this
task: We explain the main concepts related to the ray transform, its
kernel, and the extent to which it can be inverted, for the special
case of vector- and symmetric tensor fields in $\RR^3$. Our goal is on
the one hand to provide an overview of the mathematical structure of
the ray transform, and on the other hand to illustrate the method by
discussing a simple application within the field of Integrated
Photoelasticity. With these objectives in mind we have performed a new
proof on one of the central theorems in Sharafutdinov's theory, namely
the kernel of the ray transform; this theorem provides essential
insight into the degree to which optical or stress tensors can be
reconstructed from photoelastic data. Our proof focuses on vector- and
symmetric $2$-tensors only, which moreover are reconstructed
plane-wise, so that they are effectively defined on a two-dimensional
$\RR^2$. This allows us to take alternative routes in the various
steps of the proof which are not available in the most general
case. This proof is given along the way of explaining the emergence of
the ray transform in photoelastic stress reconstruction.

The plan of the paper is as follows: In section~\ref{BasicEquations}
we derive the basic equations of Integrated Photoelasticity governing
the evolution of the components of the electric field vector of a
polarized light beam passing through a medium with spatially varying
dielectric tensor. In section~\ref{StressOptRelations} we show how
these equations can be used to describe the impact of stress in a
transparent material on polarized light. In section~\ref{RayTransform}
we show how the concept of the ray transform emerges as we try to
reconstruct optical and stress tensors from photoelastic measurement
data. In section~\ref{KernelRT} we characterize the kernel of the ray
transform of vector and symmetric tensor fields, providing a new
proof, albeit with a narrower scope, to Sharafutdinov's general
theorem. In section~\ref{LongTransComponents} we show how the kernel
of the ray transform is related to the decomposition of tensor fields
into longitudinal and transverse components and explain how this
knowledge can be utilized to reconstruct stress tensor components in a
photoelastic specimen. Finally, in section~\ref{Kantarovich} we
speculate about the extension of the formalism at hand into the
high-stress regime using the Newton-Kantarovich method. In
section~\ref{summary} we summarize our results.

\section{Basic equations of Integrated Photoelasticity}
\label{BasicEquations}

We now derive the basic equations of Integrated Photoelasticity under
the {\it straight-line assumption}, i.e., light rays passing through
photoelastic media are assumed to propagate along straight lines. This
assumption is justified as long as anisotropy is sufficiently weak.

We wish to examine the propagation of a plane polarized
electromagnetic wave through a weakly birefringent photoelastic
medium. The birefringence, i.e., the anisotropy of the dielectric
tensor $\epsilon_{ij}, i,j =1,2,3$, is assumed to be entirely
resulting from the stress within the material, such that, in the
unloaded state, the material is homogeneous and isotropic with a real
dielectric tensor $\epsilon_{ij} = \epsilon\, \delta_{ij}$, and
$\epsilon = const$. The material is assumed to be unmagnetic with
vacuum permeability $\mu_0$, and non-absorbing at least in the optical
range of wavelengths, so that the material is transparent to visible
light. The permittivity of vacuum will be denoted as $\epsilon_0$.

We start with the source-free Maxwell equations in the dielectric
medium,
\begin{subequation}
\label{bas1}
\begin{eqnarray}
 \na \times \v{H} = \dot{\v{D}} \quad & , & \quad \na \times \v{E} = -
 \dot{\v{B}} \quad, \label{bas1a} \\
 \na \bl \v{E} = 0 \quad & , & \quad \na \bl \v{B} = 0 \quad,
 \label{bas1b}
\end{eqnarray}
\end{subequation}
where we have ignored the microscopic sources of the dielectric; they
are taken into account phenomenologically by the spatially varying
dielectric tensor $\epsilon_{ij}$, which links the electric field
$\v{E}$ and the displacement field $\v{D}$ according to
\begin{equation}
\label{link1}
 D_i = \epsilon_{ij}\, E_j \quad,
\end{equation}
where the convention of summing over double indices is adopted.
Furthermore, the effective charge density due to polarization effects
is ignored,
\begin{equation}
\label{bas0}
\rho_{{\rm eff}} = - \na \bl \v{P} = 0 \quad.
\end{equation}
If the second equation in (\ref{bas1a}) is inserted into the first, we
obtain the wave equation
\begin{equation}
\label{bas2}
-\Delta \v{E} + \mu_0 \, \ddot{\v{D}} = 0 \quad,
\end{equation}
where $\Delta$ is the three-dimensional Laplace operator in Cartesian
coordinates.

We now assume a harmonic time dependence $\sim e^{-i\omega t}$ of all
fields, and a wave propagation in the $z$-direction. We may neglect
the $x$- and $y$-dependence of the fields, since the electromagnetic
energy in the geometrical-optics limit propagates along narrow tubes
of light rays in such a way that, along the cross-section of a given
tube, the fields may be treated as independent of $x$ and $y$. Then
(\ref{bas2}) becomes a Helmholtz equation
\begin{equation}
\label{bas3}
\frac{d \v{E}}{d z^2} + \mu \omega^2 \, \v{D} = 0 \quad.
\end{equation}
Provided that birefringence is weak, i.e., the load on the material is
not too strong, the component of the electric field in the direction
$\v{e}_z$ of the wave vector can be neglected. This leads to an Ansatz
for the electric field as follows:
\begin{subequation}
\label{bas4}
\begin{eqnarray}
 \v{E}(z,t) & = & \mathfrak{Re} \left\{ A_1(z) \, \v{e}_1 + A_2(z) \,
 \v{e}_2 \right\} \, e^{ikz-i\omega t} \quad, \label{bas4a} \\
 k & = & \frac{\omega}{u} \quad, \quad u = \frac{1}{\sqrt{\mu_0
 \epsilon}} \quad. \label{bas4b}
\end{eqnarray}
\end{subequation}
This Ansatz contains the assumption that the phase velocity $u$ of the
wave is homogeneous and isotropic, and is determined by the
permittivity $\epsilon$ of the unstressed medium; this is true under
the same conditions under which the $z$-component of the electric
field can be neglected, i.e. sufficiently weak
anisotropy. Eqs. (\ref{bas4}) express the {\it straight-line
assumption} mentioned at the beginning of this section.

Now the the second derivative in $z$ of the complex electric field in
(\ref{bas4}) is
\begin{equation}
\label{bas5}
 E''_i = A''_i \, e^{ikz} + 2ik \, A'_i \, e^{ikz} - k^2 \, A_i \,
 e^{ikz} \quad, \quad i=1,2 \quad,
\end{equation}
where a prime indicates a derivative $\d/\d z$. If the medium were
homogeneous, the amplitudes $A_i$ were independent of $z$, just as
within vacuum. The inhomogeneity of the medium makes them
$z$-dependent, but as long as birefringence is weak, this dependence
is sufficiently weak to neglect the second derivative $A''_i$ in
(\ref{bas5}). If we then insert (\ref{bas5}) into (\ref{bas3}) and
divide by $2ik$, we obtain
\begin{equation}
\label{bas6}
 A'_i = \frac{k}{2i} \, A_i - \frac{k}{2i\epsilon} \, \epsilon_{ij} \,
 A_j \quad, \quad i=1,2 \quad.
\end{equation}
We introduce a constant
\begin{equation}
\label{bas7}
 C_0 = \frac{k}{2\epsilon} = \frac{\omega}{2c\sqrt{\epsilon_0
 \epsilon}} \quad,
\end{equation}
with the help of which (\ref{bas6}) can be written as
\begin{subequation}
\label{DynEquations}
\begin{equation}
\label{bas8}
 \frac{d A_i}{d z} = A'_i = i \, C_0 \, \left( \epsilon_{ij} \, A_j -
 \epsilon \, A_i \right) \quad, \quad i=1,2 \quad,
\end{equation}
or
as a matrix equation
\begin{equation}
\label{bas10}
 \frac{d}{d z} \left( \begin{mat}{1} &A_1\\ &A_2 \end{mat} \right) =
 i\, M(z) \left( \begin{mat}{1} &A_1\\ &A_2 \end{mat} \right) \quad,
 \quad M(z) = C_0 \left( \begin{mat}{2} &\epsilon_{11}-\epsilon & &
 \epsilon_{12} \\ &\epsilon_{21} & & \epsilon_{22}-\epsilon \end{mat}
 \right) \quad.
\end{equation}
\end{subequation}
The complex two-component vector $A \equiv (A_1,A_2)^T$ is called the
{\it Jones vector} of the light beam. Eq. (\ref{bas10}) can be
converted into an operator equation as follows: We may partition the
photoelastic medium into many thin slices perpendicular to the
direction of the wave propagation $\v{e}_z$. Each slice acts on the
Jones vector of the light beam either as a {\it retarder}, i.e.,
introducing a phase shift between two orthogonally polarized
components of the wave, or as a {\it rotator}, i.e., rotating the
plane of polarization of a plane-polarized wave, or both. This implies
that each slice acts on the local Jones vector as a unitary matrix,
and hence the same must be true for the total bulk of the photoelastic
specimen. If $z_i$ is some point before the medium (possibly
$z_i=-\infty$) and $z_f$ is some point behind the medium (possibly
$z_f=+\infty$), then the Jones vectors $A_i$ and $A_f$ at these points
must be related by a unitary transformation $U(z_f,z_i)$. In fact,
such a transformation must exist for any pair of points $z, z'$, so
that we can write
\begin{equation}
\label{ope1}
 A_{z} = U(z,z') \; A_{z'} \quad {\rm for~ all ~} z, z' \ms.
\end{equation}
The matrix $U(z,z')$ must satisfy an equation analogous to
(\ref{bas10}),
\begin{equation}
\label{ope2}
\frac{d}{d z} \, U(z,z') = i \, M(z) \; U(z,z') \quad.
\end{equation}
The matrices $U(z,z')$ obviously must have the group property,
\linebreak
\parbox{6em}{
\begin{eqnarray*}
 \hspace{5em} U(z_3,z_1) & = & U(z_3,z_2) \; U(z_2,z_1) \quad, \\
 \hspace{5em} U(z_1,z_2) & = & U(z_2,z_1)^{-1} \quad, \quad U(z,z) = \Eins_2 \quad.
\end{eqnarray*}
} \hfill
\parbox{2em}{\begin{eqnarray} \label{ope3} \end{eqnarray}}

\section{Stress-optical relations} \label{StressOptRelations}

The {\it stress-optical equations} linking the dielectric tensor
$\epsilon_{ij}$ with the stress tensor $\sigma_{ij}$,
\begin{equation}
\label{stress1}
 \epsilon_{ij} = \epsilon \, \delta_{ij} + C_1 \, \sigma_{ij} + C_2 \,
 \tr{\sigma} \, \delta_{ij} \quad,
\end{equation}
were discovered long ago by Maxwell \cite{MaxwellStrOpt}. Since the
material is assumed homogeneous with respect to its stress-optical
properties, the quantities $C_1$ and $C_2$ are constants,
characterising the stress-optical properties of the specimen. In
(\ref{stress1}), $\epsilon_{ij}$ and $\sigma_{ij}$ depend on the
spatial coordinates $(x,y,z)$, but $\epsilon$ is constant throughout
the material, as mentioned in section \ref{BasicEquations}. Using
(\ref{stress1}) we can express the equations (\ref{DynEquations}) in
terms of the stress tensor $\sigma_{ij}$ rather than the dielectric
tensor $\epsilon_{ij}$. We start again with (\ref{bas10}) and
substitute (\ref{stress1}) for the components of the dielectric tensor
in the matrix $M(z)$; this gives
\begin{equation}
\label{str2}
 M(z) = C_0 \left( \begin{mat}{2} &C_1\, \sigma_{11} + C_2\,
 \tr{\sigma} & & C_1\, \sigma_{12} \\ &C_1\, \sigma_{21} & & C_1\,
 \sigma_{22} + C_2\, \tr{\sigma} \end{mat} \right) \quad,
\end{equation}
where the trace $\tr{\sigma}$ runs over all three indices,
\begin{equation}
\label{str3}
 \tr{\sigma}  \equiv \sum\limits_{n=1}^3 \sigma_{nn} \quad.
\end{equation}
We now perform the split
\begin{subequation}
\label{str4}
\begin{eqnarray}
\label{str4a}
 C_1\, \sigma_{11} + C_2\, \tr{\sigma} & = & \ms \frac{1}{2}\, C_1\,
 \left( \sigma_{11} - \sigma_{22} \right) + B \quad, \\
 C_1\, \sigma_{22} + C_2\, \tr{\sigma} & = & - \frac{1}{2}\, C_1\,
 \left( \sigma_{11} - \sigma_{22} \right) + B \quad,
\end{eqnarray}
where
\begin{equation}
\label{str4b}
 B = \frac{1}{2}\, C_1\, (\sigma_{11} + \sigma_{22}) + C_2\,
 \tr{\sigma} \quad.
\end{equation}
\end{subequation}
When inserted into (\ref{str2}), the matrix $M$ can be written as
\begin{subequation}
\label{str5}
\begin{eqnarray}
 M(z) & = & C_0\, C_1\, \left[ \begin{mat}{2} & \frac{1}{2}
 (\sigma_{11} - \sigma_{22}) & & \sigma_{12} \\ &\sigma_{21} & -&
 \frac{1}{2} (\sigma_{11} - \sigma_{22}) \end{mat} \right] + C_0\, B\,
 \Eins_2 = \label{str5a} \\[5pt]
 & = & N(z)\, \quad +\quad C_0\, B\, \Eins_2 \quad. \label{str5b}
\end{eqnarray}
\end{subequation}
We see that the term $C_0B$ just introduces a phase common to both
components $A_1$ and $A_2$ of the Jones vector. This phase is
physically immaterial and can be absorbed by defining a "reduced"
Jones vector $A_r$,
\begin{equation}
\label{str6Cor}
 A_r \equiv A\; \exp\left( -i C_0 \int\limits_{-\infty}^z dz'\; B(z')
 \right) \quad.
\end{equation}
The dynamical equations (\ref{DynEquations}) expressed in terms of
$A_r$ then take the simpler form
\begin{subequation}
\label{str7}
\begin{eqnarray}
 \frac{d}{dz}\, A_r(z) & = & i\, N(z)\, A_r(z) \quad,
 \label{str7a} \\[10pt]
 N(z) & = & \frac{1}{2}\, C_0\, C_1\, \left[
 \begin{mat}{2} & \sigma_{11} - \sigma_{22} & & 2\, \sigma_{12} \\
 &2\, \sigma_{21} & -& \left( \sigma_{11} - \sigma_{22} \right)
 \end{mat} \right] \quad. \label{str7b}
\end{eqnarray}
\end{subequation}
Since Jones vectors at different points are unitarily related, see
(\ref{ope1}), eq. (\ref{str7a}) gives rise to an operator differential
equation analogous to (\ref{bas10}),
\begin{equation}
\label{opeq1}
\frac{d}{d z} \, U(z,z') = i \, N(z) \; U(z,z') \quad, \quad U(z,z) =
\Eins_2 \quad.
\end{equation}
This can be converted into an integral equation
\begin{equation}
\label{opeq2}
 U(z,z') = \Eins_2 + i \int\limits_{z'}^z dz_1\; N(z_1)\, U(z_1,z')
 \quad.
\end{equation}
A formal solution is given by the {\it Born-Neumann series}
\begin{equation}
\label{BornNeumann1}
 U(z,z') = \Eins_2 + \int\limits_{z'}^z dz_1\; i\, N(z_1) +
 \int\limits_{z'}^z dz_1\; i\, N(z_1) \int\limits_{z'}^{z_1} dz_2\;
 i\, N(z_2) + \cdots \quad.
\end{equation}
If the specimen is located such that the $z$-axis intersects the
medium, and $z,z'$ refer to points before and behind the medium,
respectively, then the matrix $U(z,z')$ contains all the data acquired
from measurements of the phase retardation between orthogonal
components of plane-polarized light, and the rotation of the plane of
polarized light, along that line of intersection.

For weak birefringence, (\ref{BornNeumann1}) can be truncated after
the first-order term,
\begin{equation}
\label{BN2}
  U(z,z') = \Eins_2 + i\int\limits_{z'}^{z} dz_1\; N(z_1) \quad.
\end{equation}
Since the medium occupies a bounded region in space, the limits $z$
and $z'$ in the integral in (\ref{BN2}) may be taken as $\pm
\infty$. On inserting (\ref{str7b}) into (\ref{BN2}) we see that the
evolution of the state of polarization along the light ray is
determined by the two integrals
\begin{subequation}
\label{rec6}
\begin{eqnarray}
 Iw & = & \int\limits_{-\infty}^{\infty} dz\; \sigma_{12} \quad, \quad
 \\
 Iu & = & \int\limits_{-\infty}^{\infty} dz\; (\sigma_{11}-
 \sigma_{22}) \quad.
\end{eqnarray}
\end{subequation}
The significance of the notation $Iw, Iu$ will become clear shortly.

\section{The ray transform in Photoelasticity} \label{RayTransform}

We now wish to examine the stress within a cylindrical object which
occupies a cylindrical domain $G=D \times (a,b)$, where $D$ is a
two-dimensional region in the $xy$-plane and the boundary $\d D$ of
$D$ is a strictly convex smooth curve. By $B = \d D \times (a,b)$ we
denote the lateral surface of the cylinder $G$. The stress tensor is
supposed to be smooth in $G$ and on $B$, and satisfies the {\it
equilibrium conditions}
\begin{equation}
\label{rec8}
 \frac{\d}{\d x_j}\, \sigma_{ij} = 0 \quad
\end{equation}
everywhere. Furthermore, the absence of external forces on the lateral
boundary $\d B$ implies that
\begin{equation}
\label{rec9}
 \sigma_{ij}\, n_j = 0 \quad,
\end{equation}
where $(n_1,n_2,0)$ is the unit outward normal to $\d B$ on the
lateral surface $B$ of the cylinder. Strain is therefore applied only
at the top ($z=b$) and the bottom ($z=a$) of the object.

It is now assumed that measurements analogous to (\ref{BN2}) are
performed along all horizontal straight lines in $\RR^3$ defined by $t
\mapsto X + t\, \xi(\alpha)$, where $X=(x,y,z)$, $(x,y)$ varies
through the points in the $xy$-plane, $z$ takes values in the interval
$(a,b)$, the angle $\alpha$ varies in $(0,2\pi)$, and
\begin{equation}
\label{defxi}
 \xi(\alpha) = \big( \cos\alpha, \sin \alpha, 0 \big) \quad
\end{equation}
is the horizontal unit vector in the direction of the line. These
lines generalize the measurements along the $z$-axis which were
discussed in section \ref{BasicEquations}. For the straight line
passing through the point $X$ along the direction $\xi$ we now choose
an adapted right-handed coordinate system $(\eta,z,t)$ in which the
direction of propagation of the light beam is along the positive
$t$-axis (the $\eta$-coordinate line is perpendicular to the
$t$-coordinate line, and the new $z$-direction coincides with the old
one). On performing a measurement along the line so defined we then
obtain quantities analogous to (\ref{rec6}),
\begin{subequation}
\label{rec10}
\begin{eqnarray}
 Iw(X, \xi) & = & \int\limits_{-\infty}^{\infty} dt\; \sigma_{\eta z}
 \quad, \quad \label{rec10a} \\
 Iu(X, \xi) & = & \int\limits_{-\infty}^{\infty} dt\;
 (\sigma_{\eta\eta}- \sigma_{zz}) \quad, \label{rec10b}
\end{eqnarray}
\end{subequation}
the notation on the left-hand side indicating that these lines pass
through the point $X$ and are directed along the unit vector
$\xi$. The tensor components in these integrals can be expressed in
terms of the laboratory-frame components $\sigma_{ij}$ as
\begin{subequation}
\label{rec13}
\begin{eqnarray}
 \left( \sigma_{\eta\eta} - \sigma_{zz} \right)(t,0,0) & = &
 \sin^2\alpha\; (\sigma_{11} - \sigma_{33})(X+t\xi) - \nonumber \\
 & - & 2\, \sin\alpha\, \cos\alpha\; \sigma_{12}(X+t\xi) + \nonumber
 \\
 & + & \cos^2\alpha\; (\sigma_{22} - \sigma_{33})(X+t\xi) \quad,
 \label{rec13a} \\
 \sigma_{\eta z}(t,0,0) & = & -\sin\alpha\; \sigma_{13}(X+t\xi) +
 \nonumber \\
 & + & \cos\alpha\; \sigma_{23}(X+t \xi) \quad, \label{rec13b} \\
 \sigma{zz}(t,0,0) & = & \sigma_{33}(X+t\xi) \quad. \label{rec13c}
\end{eqnarray}
\end{subequation}
We now define a vector field $w$ in $\RR^2_z$, where
\begin{equation}
\label{r2z1}
 \RR^2_z \equiv \left\{ (x,y,z) \in \RR^3 \right|\left. \, (x,y) \in
\RR^2\, , \; z \in (a,b) \; \rm{fixed} \right\} \quad,
\end{equation}
by
\begin{subequation}
\label{tensors1}
\begin{equation}
\label{vec1}
 \left( w_1, w_2 \right) = \left( \sigma_{23}, -\sigma_{13} \right)
 \quad,
\end{equation}
and a symmetric tensor field $u$ by
\begin{equation}
\label{tens1}
 \left(u_{11}, u_{12}, u_{21}, u_{22} \right) = \left(
 \sigma_{22}-\sigma_{33}, -\sigma_{12}, -\sigma_{21}, \sigma_{11} -
 \sigma_{33} \right) \quad.
\end{equation}
\end{subequation}
If (\ref{rec13}) is expressed in terms of the components of the
tensors $u$ and $w$ and inserted into (\ref{rec10}) we obtain
\begin{subequation}
\label{rec14}
\begin{eqnarray}
 Iw(X,\xi) & = & \int\limits dt\; w_m\, \xi_m \quad, \label{rec14a}
 \quad \\
 Iu(X,\xi) & = & \int\limits dt\; u_{mn}\, \xi_m\, \xi_n
 \quad. \label{rec14b}
\end{eqnarray}
\end{subequation}
According to \cite{Sharafutdinov}, the collection of quantities
$Iw(X,\xi)$, taken for all $X \in \RR^2_z$ and $\xi=\xi(\alpha)$ with
$\alpha \in [0,2\pi]$, constitutes the {\it ray transform} of the
two-dimensional vector field $w$; whilst the collection of all
$Iu(X,\xi)$, with the same range of variables, constitutes the ray
transform for the two-dimensional symmetric tensor $u$.

The general definition of the ray transform of a symmetric tensor
field $T_{i_1 \ldots i_m}$ of degree $m$ defined in $\RR^n$, which
includes both $Iw$ and $Iu$ in (\ref{rec14}) as special cases, is
given by
\begin{equation}
\label{rec15}
 IT(X,\xi) = \int dt\; T_{i_1\ldots i_m}(X +
 t\xi)\, \xi_{i_1}\, \ldots\, \xi_{i_m} \quad,
\end{equation}
where $X, \xi \in \RR^n$ with $|\xi|=1$, i.e. $\xi$ lies on the unit
sphere $S_{n-1}$ in $\RR^n$. Since the ray transform is constant for
all $X'$ lying on the line $t\mapsto X + t\, \xi$,
\begin{equation}
\label{con1}
 IT(X+\tau\, \xi, \xi) = IT(X, \xi) \quad,
\end{equation}
we can restrict the range of the variable $X$ to the subspace
\begin{equation}
\label{con2}
 \xi_{\bot} = \big\{ x\in\RR^n \big| \; x \bot \xi \;\big\}
\end{equation}
of $\RR^n$ orthogonal to $\xi$, since any component of $X$ in the
direction of $\xi$ may be set to zero. It follows then that the ray
transform is really a function on the {\it tangent bundle} $TS_{n-1}$
of the unit sphere $S_{n-1}$,
\begin{equation}
\label{con3}
 TS_{n-1} = \big\{ (X,\xi) \in \RR^n \times S_{n-1} \big|\; X \bot
 \xi\; \big\} \quad.
\end{equation}
As a consequence, the Fourier transform with respect to $X$ may be
restricted to the subspace $\xi_{\bot}$,
\begin{equation}
\label{rec17}
 \widehat{IT}(k,\xi) = \frac{1}{(2\pi)^{(n-1)/2}}
 \int\limits_{\xi^{\bot}} dV^{n-1}(X')\; IT(X',\xi)\, e^{-i\bra k,
 X'\ket} \quad, \quad k \in \xi_{\bot} \quad.
\end{equation}

The Fourier transform (\ref{rec17}) of the ray transform is related to
the Fourier transform $\widehat{T}$ of the tensor $T$ as follows:
\begin{equation}
\label{con4}
 \widehat{IT}(k,\xi) = (2\pi)^{1/2}\; \widehat{T}_{i_1\ldots i_m}(k)\,
 \xi_{i_1} \ldots \xi_{i_m} \quad, \quad k\in \xi_{\bot} \quad.
\end{equation}
This can be proven easily: Inserting (\ref{rec15}) into (\ref{rec17})
we find
\begin{equation}
\label{con5}
 \widehat{IT}(k,\xi) = \frac{1}{(2\pi)^{(n-1)/2}}
 \int\limits_{\xi_{\bot}} dV^{n-1}(X') \int dt\; T_{i_1 \ldots i_m}\,
 \xi_{i_1} \ldots \xi_{i_m}\; e^{-i\bra k,X'\ket} \quad.
\end{equation}
Since $|\xi|=1$ and $X'\bot\xi$, we can introduce coordinates $X$ on
$\RR^n$ by $X \equiv (X',t)$, in which case $dV^{n-1}(X')\, dt =
dV^n(X)$. Furthermore, since $k \bot \xi$ by definition (\ref{rec17})
of the Fourier transform, the argument of the exponential is
\begin{equation}
\label{con6}
 \bra k, X \ket = \bra k, X'+t\, \xi \ket = \bra k, X' \ket \quad.
\end{equation}
If these relations are inserted into (\ref{con5}), (\ref{con4})
follows.

\section{The kernel of the ray transform} \label{KernelRT}

Our task is now to reconstruct the components of the stress tensor
from the ray transforms (\ref{rec14}) as far as possible. We shall see
that, for the given experimental setup, involving horizontal light
rays only, we can only retrieve one component of the stress tensor,
namely $\sigma_{33}$. In order to proceed we need to understand that
the ray transform, regarded as an operator $I$ sending symmetric
tensor fields $T$ (including fields of degree $m=1$, i.e. vector
fields) $T$ to the set of quantities $IT$, has a nontrivial kernel:
Given two tensor fields $T^{(1)}_{i_1 \ldots i_m}$ and $T^{(2)}_{i_1
\ldots i_m}$, does $IT^{(1)} = IT^{(2)}$ imply $T^{(1)} = T^{(2)}$?
The answer is no, as proved in a theorem in
{\cite{Sharafutdinov}}. The original version of this theorem is
designed to encompass the most general case of symmetric $m$-tensors
defined on $\RR^n$, with no reference to a specific physical
application or context. Accordingly, the proof, encumbered with this
degree of generality, occupies a substantial number of pages. However,
this generality is not required for the particular purposes of
reconstructing vector- and symmetric $2$-tensor fields within the
context of Photoelasticity. If we limit our scope to these types of
fields, the proof of the theorem can be performed along alternative
routes which are shorter than in the original work. In this section we
provide this new proof for the two special cases of a vector field $w$
and a symmetric tensor field $u$ defined in $\RR^2_z$ as discussed in
(\ref{tensors1}, \ref{rec14}):
\begin{theorem}[Kernel of ray transform] \label{kernelTheorem}
\begin{subequation}
\begin{description}
\item[A.] \label{Vec}
Let $w$ be a smooth vector field in $\RR^2$ with compact support. Then
the following statements are equivalent:
\begin{description}
\item[a)]
 Iw = 0 \quad. \label{kerVeca}
\item[b)]
There exists a compactly supported scalar field $\phi$ on $\RR^2$ such
that its support is contained in the convex hull of the support of
$w$, and
\begin{equation}
\label{kerVecb}
 w_i = \d_i \phi \quad, \quad i=1,2 \quad.
\end{equation}
\item[c)]
The identity
\begin{equation}
\label{SVa}
 \d_1 w_2 - \d_2 w_1 = 0 \quad
\end{equation}  holds.
\end{description}
\item[B.] \label{Ten}
Let $u$ be a symmetric tensor field on $\RR^2$ with compact
support. Then the following statements are equivalent:
\begin{description}
\item[a)]
 Iu = 0 \quad. \label{kerTena}
\item[b)]
There exists a compactly supported vector field $v$ in $\RR^2$ such
that its support is contained in the convex hull of the support of
$u$, and
\begin{equation}
\label{kerTenb}
 u_{ij} = \frac{1}{2} \left( \d_i v_j + \d_j v_i \right) \quad, \quad
 i=1,2 \quad.
\end{equation}
\item[c)]
The identity
\begin{equation}
\label{SVb}
 \d_1^2\, u_{22} + \d_2^2\, u_{11} - 2\, \d_1 \d_2\, u_{12} = 0 \quad,
\end{equation}
holds.
\end{description}
\end{description}
\end{subequation}
\end{theorem}

\vspace{1em}

\begin{pf}

\subsubsection{Case A.}

\vspace{0.5em}

The equivalence (A.b) $\Leftrightarrow$ (A.c) is well-known in
standard Vector Analysis, where (\ref{SVa}) expresses the fact that
the vector field $w_i$ must be curl-free in order for a potential
$\phi$ to exist. We therefore shall not prove this equivalence here.

\vspace{0.5em}

We now prove (A.b) $\Rightarrow$ (A.a): Inserting (\ref{kerVecb}) into
the first integral of (\ref{rec14}) yields
\begin{equation}
\label{proA1}
 Iw(X,\xi) = \int dt\; \xi_m \d_m \phi(X+t\,\xi) = \int dt\;
 \frac{d}{dt}\, \phi(X+t\,\xi) = 0 \quad,
\end{equation}
since $\phi$ has compact support and therefore must vanish at $t= \pm
\infty$.

\vspace{0.5em}

Finally we prove (A.a) $\Rightarrow$ (A.c): To this end we observe
that the function $\d_1 w_2 - \d_2 w_1$ is a scalar with respect to
rotations in the plane $\RR^2$. This is obvious if we extend the field
$w_i, i=1,2$, smoothly to a three-dimensional vector field $\bar{w}_i,
i=1,2,3$, defined in $\RR^3$, such that $\bar{w}_1 = w_1$ and
$\bar{w}_2 = w_2$ on the plane $z= const$. Then (\ref{SVa}) is just
the $z$-component of the curl of $\hat{w}$, which transforms like a
scalar under $SO(2)$-rotations in this plane. It follows that the
collection of integrals
\begin{equation}
\label{radon1}
 \int dt\; \big( \d_1 w_2 - \d_2 w_1 \big)(X+t\xi) \quad,
\end{equation}
taken for all directions $\xi$ in $\RR^2$, is just the two-dimensional
{\it Radon transform} of this scalar field. If all of the integrals
(\ref{radon1}) were zero, the invertibility of the Radon transform
would imply that the integrand must vanish and (\ref{SVa}) must
hold. But this is indeed the case: Let us take the derivative of the
equation $Iw=0$ with respect to $x_1$,
\begin{equation}
\label{radon2}
 \d_1\, Iw = \int dt\; \big\{ \cos\theta\, w_{1,1} + \sin\theta\,
 w_{2,1} \big\} = 0 \quad,
\end{equation}
where we use the abbreviation $w_{1,1} \equiv \d_1 w_1$, etc. On
account of the fact that $w_1$ has compact support and therefore must
vanish at infinity we have
\begin{equation}
\label{radon3}
 \int dt\; \big\{ \cos\theta\, w_{1,1} + \sin\theta\, w_{1,2} \big\} =
 \int dt\; \frac{d}{dt} w_1 = 0 \quad,
\end{equation}
so that (\ref{radon2}) gives
\begin{equation}
\label{radon4}
 \d_1 Iw = \sin\theta \int dt \big\{ w_{2,1} - w_{1,2} \big\} = 0
 \quad.
\end{equation}
From $\d_2 Iw =0$ we infer a similar equation, with $\sin\theta$
replaced by $\cos\theta$ in front of the integral. These equations
must hold for all $\theta$, so that the line integrals (\ref{radon1})
all vanish. This proves the statement.

\vspace{0.5em}

This finishes the proof of case (A.).

\subsubsection{Case B.}

\vspace{0.5em}

Our strategy here is different, since we cannot employ standard
techniques from Vector Analysis.

\vspace{0.5em}

We first prove the equivalence (B.b) $\Leftrightarrow$ (B.c): The
implication (B.b) $\Rightarrow$ (B.c) is trivial, and follows
immediately by inserting (\ref{kerTenb}) into (\ref{SVb}).

\vspace{0.5em}

Now we prove (B.c) $\Rightarrow$ (B.b): To this end we define a
two-dimensional vector field $(v_1, v_2)$ as follows:
\begin{eqnarray}
\label{potent1}
 v_i & = & \frac{1}{\pi} \int\limits_0^{2\pi} d\theta
 \int\limits_{-\infty}^0 dt\; u_{mn}(X+ t\, \xi)\; \xi_m \xi_n \xi_i
 \quad,
\end{eqnarray}
where $i=1,2$, $\xi_1 = \cos\theta,\; \xi_2 = \sin\theta$. We can now
compute the expression $\d_1 \d_1 v_1 \equiv v_{1,11}$, by using the
fact that
\begin{equation}
\label{potent2}
 \frac{d}{dt}\, u_{11} = \cos\theta\, u_{11,1} + \sin\theta\, u_{11,2}
 \quad.
\end{equation}
Making use of the assumption (\ref{SVb}) we are able to derive
\begin{subequation}
\label{potent3}
\begin{equation}
\label{potent3a}
 v_{1,11} = u_{11,1} \quad.
\end{equation}
In a similar way we find
\begin{equation}
\label{potent3b}
 v_{1,12} = u_{11,2} \quad.
\end{equation}
\end{subequation}
Eqs. (\ref{potent3}) imply that $\d_1 v_1 = u_{11} + c$, where $c$ is
a constant. However, $u_{11}$ has compact support and therefore
vanishes at infinity, whilst $v_1$ by construction behaves like
$1/|X|$ for $|X| \rightarrow \infty$ and, in particular, vanishes at
infinity . It follows that $c=0$, and hence
\begin{equation}
 u_{11} = \d_1 v_1 \quad,
\end{equation}
which proves eq. (\ref{kerTenb}) for the case $i=j=1$. The proof for
$u_{12}$ and $u_{22}$ proceeds in exactly analogous a manner.

\vspace{0.5em}

Next we prove (B.b) $\Rightarrow$ (B.a): Inserting (\ref{kerTenb})
into (\ref{rec14b}) yields
\begin{equation}
\label{proB1}
 Iu(X,\xi) = \int dt\; \xi_m \xi_n \d_m v_n(X+t\,\xi) = \int dt\;
 \xi_n\, \frac{d}{dt}\, v_n(X+t\,\xi) = 0 \quad,
\end{equation}
since $v$ has compact support.

\vspace{0.5em}

Finally we prove (B.c) $\Rightarrow$ (B.a): Using the definition
(\ref{rec14b}) of $Iu$, we derive the equations
\begin{subequation}
\label{proB6}
\begin{eqnarray}
 (Iu)_{,11} & = & \int dt\; \bigg\{\, \cos^2\theta\, u_{11,11} +
 \nonumber\\
 & + & 2\sin\theta \cos\theta\, u_{12,11} + \sin^2\theta\, u_{22,11}
 \bigg\} \quad, \label{proB6a} \\
 (Iu)_{,12} & = & \int dt\; \big\{\, \cos^2\theta\, u_{11,12} +
 \nonumber \\
 & + & 2\sin\theta \cos\theta\, u_{12,12} + \sin^2\theta\, u_{22,12}
 \bigg\} \quad, \label{proB6b} \\
 (Iu)_{,22} & = & \int dt\; \big\{\, \cos^2\theta\, u_{11,22} +
 \nonumber\\
 & + & 2\sin\theta \cos\theta\, u_{12,22} + \sin^2\theta\, u_{22,22}
 \bigg\} \quad. \label{proB6c}
\end{eqnarray}
\end{subequation}
Starting with (\ref{proB6b}) we find on using (\ref{SVb}) that
\begin{equation}
\label{proB7}
 (Iu)_{,12} = \int dt\; \bigg\{\, \cos\theta\, \frac{d}{dt}\, u_{11,2}
 + \sin\theta\, \frac{d}{dt}\, u_{22,1}\, \bigg\} = 0 \quad.
\end{equation}
Using the same technique we obtain for (\ref{proB6a}, \ref{proB6c})
\begin{equation}
\label{proB8}
 (Iu)_{,11} = (Iu)_{,22} = 0 \quad.
\end{equation}
It follows that
\begin{equation}
\label{proB10}
 (Iu)_{,1} = f_1(\theta) \quad, \quad (Iu)_{,2} = f_2(\theta) \quad.
\end{equation}
The functions $f_1$ and $f_2$ can be evaluated at points $X$
arbitrarily close to infinity, so using the fact that $u$ has compact
support implies that $f_1 = f_2 = 0$. Then, repeating the same
argument for $Iu_{,1}$ and $Iu_{,2}$ shows that we must have $Iu=0$.

\vspace{0.5em}

This finishes the proof of case (B.).

\vspace{0.5em}

This finishes the proof of theorem \ref{kernelTheorem}. \hfill
$\blacksquare$
\end{pf}

From eqs. (\ref{SVa}) and (\ref{SVb}) we learn that elements of the
kernel of the ray transform can be characterized by the identical
vanishing of a differential expression of the vector/tensor components
$w_i, u_{ij}$. The differential operator acting in (\ref{SVa}) and
(\ref{SVb}) is generically called the {\it Saint-Venant operator}
\cite{Sharafutdinov}. Its vanishing can be understood as an
integrability condition ensuring the existence of the scalar/vector
fields $\phi$ and $v$ occuring in eqs. (\ref{kerVecb}) and
(\ref{kerTenb}). We see that in the simplest case of a vector field
$w$, the condition for the vanishing of the ray transform is
equivalent to the statement that $w$ is the gradient of a potential
$\phi$, and (\ref{SVa}) is nothing but the necessary and sufficient
condition for the existence of the potential. The case of a symmetric
$m$-tensor in $\RR^n$ provides a nontrivial generalization of this
scenario.

\section{Longitudinal and transverse tensor components}
\label{LongTransComponents}

The significance of the kernel of the ray transform is further
elucidated by considering the standard decomposition of symmetric
tensor fields of degree $m$ (including vector fields) into {\it
transverse} $T_{\bot i_1\ldots i_m}$ and {\it longitudinal}
$(dv)_{i_1\ldots i_m}$ components,
\begin{equation}
\label{deco}
 T_{i_1\ldots i_m} = T_{\bot i_1\ldots i_m} + (dv)_{i_1\ldots i_m}
 \quad.
\end{equation}
The transverse part $T_{\bot i_1\ldots i_m}$ is characterized by its
vanishing divergence, $\d_{i_m}\, T_{\bot i_1 \ldots i_m} = 0$, while
the longitudinal part $dv$ is the symmetrized covariant derivative of
the symmetric $(m-1)$ tensor $v$, and is given by
\begin{equation}
\label{symco1}
 \left( dv\right)_{i_1 \cdots i_m} = \frac{1}{m!} \sum\limits_{\pi \in
 \mathcal{S}_m}  v_{i_{\pi(1)} \, \ldots \, i_{\pi(m-1)}, \;
 i_{\pi(m)}} \quad,
\end{equation}
where $\mathcal{S}_m$ is the group of permutations of $m$
elements. The transverse part of $T$ is determined by the projection
of the Fourier transform $\widehat{T}_{i_1\ldots i_m}(k)$ of
$T_{i_1\ldots i_m}(X)$ onto the subspace orthogonal to the direction
of propagation $k$ of the Fourier mode $e^{i \langle k, X \rangle}$,
\begin{subequation}
\label{proj}
\begin{eqnarray}
 T_{\bot i_1\ldots i_m}(X) & = & \frac{1}{(2\pi)^{n/2}} \int d^n\!k\;
 \widehat{T}_{\bot i_1\ldots i_m}(k) \; e^{i\langle k, X\rangle}
 \quad, \label{projA} \\
 \widehat{T}_{\bot i_1\ldots i_m}(k) & = & \lambda_{i_1 j_1} \cdots
 \lambda_{i_m j_m}\; \widehat{T}_{j_1\ldots j_m}(k) \quad,
 \label{projB}
\end{eqnarray}
where
\begin{equation}
\label{projector1}
 \lambda_{ij} = \delta_{ij} - \frac{k_i k_j}{k^2} \quad
\end{equation}
\end{subequation}
is the projector onto the subspace perpendicular to $k$. By
construction, this projector leaves the Fourier transform
$\widehat{T}_{\bot}$ of the transverse component $T_{\bot}$ invariant,
while it annihilates the Fourier transform $\widehat{T} -
\widehat{T}_{\bot}$ of the longitudinal component $T- T_{\bot}$.

For our vector field $w$ and tensor field $u$, the decompositions
(\ref{deco}) take the special form
\begin{subequation}
\label{deco2}
\begin{eqnarray}
 w_i & = & w_{\bot i} + \d_i\, \phi \quad, \quad \d_i\, w_{\bot i} = 0
 \quad, \label{deco2a} \\
 u_{ij} & = & u_{\bot ij} + \frac{1}{2} \left( \d_i\, v_j + \d_j\, v_i
 \right) \quad, \quad \d_j\, u_{\bot ij} = 0 \quad. \label{deco2b}
\end{eqnarray}
\end{subequation}
On comparing (\ref{kerVecb}, \ref{kerTenb}) with (\ref{deco2a},
\ref{deco2b}) we see that the kernel of the ray transforms $Iw, Iu$
consists precisely of the longitudinal components $d\phi$, $d v$ of
the vector/tensor field $w,u$! It follows that these components cannot
be retrieved from the ray transforms of these fields. This result is
obviously of central import for any attempt to reconstruct optical and
stress tensors from photoelastic measurements.

Endowed with this knowledge we can now turn to the question to which
extent the ray transform may be inverted: To this end we introduce the
{\it integral moments} $(\mu^m IT)_{j_1 \ldots j_m}(X)$ of the ray
transform $(IT)(X,\xi)$ of the symmetric tensor $T_{i_1 \ldots
i_m}(X)$ of degree $m$ on $\RR^n$,
\begin{equation}
\label{intmo1}
 (\mu^m IT)_{j_1 \ldots j_m}(X) = \frac{1}{\omega_{n-1}}
 \int\limits_{S_{n-1}} d\Omega_{n-1}(\xi)\; \xi_{j_1} \ldots
 \xi_{j_m}\; (IT)(X,\xi) \quad,
\end{equation}
where $S_{n-1}$ is the unit sphere in $\RR^n$ with $(n-1)$-dimensional
volume $\omega_{n-1}$ and volume element $d\Omega_{n-1}(\xi)$. For our
vector field $w$ and symmetric tensor field $u$ in $\RR^2$, the
integral moments of the ray transform $Iw(X,\xi)$ are
\begin{subequation}
\label{imo}
\begin{eqnarray}
 (\mu^1 Iw)_i(X) & = & \frac{1}{2\pi} \int\limits_0^{2\pi} d\theta\;
 \xi_i\; (Iw)(X,\xi) \quad, \label{imo1} \\
 \left( \mu^2 I u \right)_{ij}(X) & = & \frac{1}{2\pi}
 \int\limits_0^{2\pi} d\theta\; \xi_i\, \xi_j\; (Iu)(X,\xi)
 \quad. \label{imo3}
\end{eqnarray}
\end{subequation}
On taking (two-dimensional) Fourier transforms of eqs. (\ref{imo1})
and (\ref{imo3}) we arrive after a lengthy calculation at
\begin{subequation}
\label{fout}
\begin{eqnarray}
 \left( \widehat{\mu^1 Iw} \right)_i(k) & = & \frac{2}{|k|}\;
 \widehat{w}_m(k)\; \lambda_{im}(k) \quad, \label{fouta} \\
 \left( \widehat{\mu^2 I u}\right)_{ij}(k) & = & \frac{2}{3} \;
 \widehat{u}_{mn}(k) \; |k|^{-1}\; \times \nonumber \\
 & \times & \bigg\{ \lambda_{ij} \lambda_{mn} + \lambda_{im}
 \lambda_{jn} + \lambda_{in} \lambda_{jm} \bigg\} \quad.
 \label{foutb}
\end{eqnarray}
\end{subequation}
The presence of the projector $\lambda_{ij}$ in these formulas cancels
out the longitudinal components of $\widehat{w}_m$ and
$\widehat{u}_{mn}$, while leaving their transverse components
$\widehat{w}_{\bot m}$ and $\widehat{u}_{\bot mn}$ invariant. It
follows that we can replace $\widehat{w}_m$ and $\widehat{u}_{mn}$ by
$\widehat{w}_{\bot m}$ and $\widehat{u}_{\bot mn}$ in formulas
(\ref{fout}). On taking the trace of the resulting equations we can
then invert them for $\widehat{w}_{\bot}$ and $\widehat{u}_{\bot}$,
\begin{subequation}
\label{inv}
\begin{eqnarray}
 \widehat{w}_{\bot i}(k) & = & \frac{|k|}{2}\, \left( \widehat{\mu^1
 Iw} \right)_i \quad, \label{inv1} \\
 \widehat{u}_{\bot ij}(k) & = & \frac{3}{4}\; |k|\; \left(
 \widehat{\mu^2 Iu} \right)_{ij} - \frac{1}{4}\; |k|\; \sum_{n=1}^2
 \left( \widehat{\mu^2 Iu} \right)_{nn}\; \lambda_{ij} \quad.
 \label{inv2}
\end{eqnarray}
\end{subequation}
Eqs. (\ref{inv}) provide the explicit answer to the question to which
degree the components of a vector field $w$ and a symmetric tensor
field $u$ on $\RR^2$ can be reconstructed from their ray transforms
$Iw$ and $Iu$: It is precisely the transverse components of these
fields which can be retrieved in full, whereas the longitudinal
components, these being symmetric covariant derivatives of lower-rank
tensors, must remain undetermined. Performing the inverse Fourier
transform on (\ref{inv}) we find that
\begin{subequation}
\label{inver10}
\begin{eqnarray}
 w_{\bot i}(X) & \equiv & \mathcal{W}_i \left[\mu^1 Iw \right](X) =
 \nonumber \\[8pt]
 & = & \frac{1}{2\pi} \int d^2\!k\; e^{i \bra k,X\ket}\;
 \frac{|k|}{2}\, \left( \widehat{\mu^1 Iw} \right)_i \quad,
 \label{inver10a} \\[12pt]
 u_{\bot ij} & \equiv & \mathcal{U}_{ij} \left[\mu^2 Iu \right](X) =
 \nonumber \\[8pt]
 & = & \frac{1}{2\pi} \int d^2\!k\; e^{i \bra k,X\ket}\; \bigg\{
 \frac{3}{4}\; |k|\; \left( \widehat{\mu^2 Iu} \right)_{ij} -
 \nonumber \\
 & - & \frac{1}{4}\; |k|\; \left( \widehat{\mu^2 Iu} \right)_{nn}\;
 \lambda_{ij} \bigg\} \quad. \label{inver10b}
\end{eqnarray}
\end{subequation}
Here, $\mathcal{W}_i \left[\mu^1 Iw \right](X)$ indicates a linear
functional with respect to $\mu^1 Iw$, being parametrized by the
spatial coordinates $X$; and similarly for $\mathcal{U}$. The
right-hand sides of (\ref{inver10}) contain the ray transforms $Iw$
and $Iu$ of the vector- and tensor field, and can therefore be
regarded as known functions of the data collected from
measurements. Now, from theorem \ref{kernelTheorem} we know that the
Saint-Vernant operator (\ref{SVa}, \ref{SVb}) annihilates precisely
the longitudinal component of a vector- or symmetric tensor field [and
therefore must be related to products of the projectors
(\ref{projector1}) by Fourier transformation]. Thus, if the
Saint-Venant operator acts on $w$ and $u$ in the manner of (\ref{SVa})
and (\ref{SVb}), only the transverse components $w_{\bot}$ and
$u_{\bot}$ of $w$ and $u$ survive. For these transverse components we
now substitute expressions (\ref{inver10}); thus, we obtain quantities
\begin{subequation}
\label{inv11}
\begin{eqnarray}
 \mathbb{W} & \equiv & \d_1\, \mathcal{W}_2 - \d_2\, \mathcal{W}_1
 \quad, \label{inv11a} \\
 \mathbb{U} & \equiv & \d_1^2\, \mathcal{U}_{22} + \d_2^2\,
 \mathcal{U}_{11} - 2\, \d_1\d_2\, \mathcal{U}_{12}
 \quad, \label{inv11b}
\end{eqnarray}
\end{subequation}
where $\mathcal{W}$ and $\mathcal{U}$ represent known functions of
measurement data. On the other hand, we can insert the values
(\ref{vec1}) and (\ref{tens1}) for the components of the vector $w$
and tensor $u$ into the left-hand sides of (\ref{inver10}), so that
\begin{subequation}
\label{retr1}
\begin{eqnarray}
 \mathbb{W} & = & - \d_1\, \sigma_{13} - \d_2\, \sigma_{23} \quad,
 \label{retr1a} \\
 \mathbb{U} & = & \d_1^2\, (\sigma_{11} - \sigma_{33}) + \d_2^2\,
 (\sigma_{22} - \sigma_{33}) + 2\, \d_1\d_2\, \sigma_{12}
 \quad. \label{retr1b}
\end{eqnarray}
\end{subequation}
We now regard (\ref{retr1}) as a system of equations for the unknown
$\sigma_{ij}$, where the left-hand sides $\mathbb{W}$ and $\mathbb{U}$
are known through (\ref{inv11}); furthermore, the equilibrium
conditions (\ref{rec8}) and the boundary conditions (\ref{rec9}) are
added to this system. The third equation in (\ref{rec8}) implies that
\begin{equation}
\label{retr2}
 \mathbb{W} = \d_3\, \sigma_{33} \quad.
\end{equation}
Furthermore, differentiating the first equation in (\ref{rec8}) with
respect to $x_1$, the second with respect to $x_2$ and adding the
results gives, on using (\ref{retr1a}),
\begin{equation}
\label{retr3}
 \d_3\, \mathbb{W} - \mathbb{U} = \left( \d^2_1 + \d^2_2 \right)\,
 \sigma_{33} \quad.
\end{equation}
This last equation defines a {\it Dirichlet problem} in $\RR^2_z$ for
the unknown $\sigma_{33}$, the left-hand side containing a known
function of measurement data.

The value of $\sigma_{33}$ on the lateral surface $B$ of the
cylindrical region $G$ is determined by the ray transform $Iu$: Let
$Iu(X,X')$ denote the ray transform of $u$ along a line which connects
two points $X=(x,y,z)$ and $X'=(x',y',z)$ at the same height $z$ on
the cylindrical specimen, where both $X$ and $X'$ belong to the
boundary $B$. By taking the limit $X' \rightarrow X$,
the integrand in $Iu$ is nonvanishing only on an infinitesimally short
interval in the neighbourhood of the point $X$, which becomes
tangential to $B$ in the limit. Using the stress-free condition
(\ref{rec9}) of the lateral surface then yields
\begin{equation}
\label{retr4}
 \lim\limits_{X' \rightarrow X} \frac{Iu(X,X')}{|X-X'|} = -
 \sigma_{33} \quad,
\end{equation}
at every point $X \in B$. Relations (\ref{retr3}, \ref{retr4}) are
sufficient to uniquely determine the component $\sigma_{33}$ from the
measured data, i.e., $\mathbb{W}$ and $\mathbb{U}$, since the value of
$\sigma_{33}$ on the boundary is given by (\ref{retr4}), hence the
Dirichlet problem (\ref{retr3}) can be solved uniquely for any given
$z \in (a,b)$. Alternatively, the Dirichlet problem may be solved for
just one $z_0 \in (a,b)$ and then (\ref{retr2}) may be integrated to
obtain $\sigma_{33}$ everywhere.

Finally, we might wonder whether there is any more information to be
retrieved from the ray transforms $Iw$ and $Iu$, or alternatively
$\mathbb{W}$ and $\mathbb{U}$, as so far we could only reconstruct one
tensor component from these data. The answer is no: As proven
in~\cite{Sharafutdinov}, if any two stress tensors $\sigma^{(1)}$ and
$\sigma^{(2)}$ satisfy (\ref{rec8}, \ref{rec9}) and $\sigma^{(1)}_{33}
= \sigma^{(2)}_{33}$ then they produce the same ray transforms,
$Iw^{(1)} = Iw^{(2)}$ and $Iu^{(1)} = Iu^{(2)}$. As a consequence, the
retrieval of one tensor component is the maximum information to be
gleaned from a ray transform which is performed {\em for horizontal
lines only}. If we wish to reconstruct other tensor components, the
orientation of the incident light beam must be changed accordingly.

\section{Newton-Kantarovich method} \label{Kantarovich}

Sharafutdinov's theory shows that in the linearised problem of
photoelastic tomography the transverse part of the stress tensor can
be recovered from complete photoelastic data, and that this inversion
is stable under suitable smoothness assumptions on $\sigma$. The
forward problem represented by the solution of the ODE (\ref{opeq1})
along each ray is non-linear in $\sigma$, its Fr\'{e}chet derivative
at $\sigma =0$ being the ray transform (\ref{BN2}). Unlike the
linearization about non-zero stress, this transform has an explicit
Fourier inversion formula. It might therefore be possible to extend
the reconstruction method outlined above into the high-stress regime
by using the Newton-Kantarovich method with fixed derivative, which
means that while the forward problem is solved successively for each
updated stress, the difference between predicted and measured optical
data is used as boundary data for the {\em same} ray transform at each
iteration. While the convergence of this method  is linear compared
with the quadratic convergence when the derivative is updated, it is
expected that the efficient computation of the linear step will make
the former more efficient numerically. These ideas will be
investigated in upcoming papers.

\section{Summary} \label{summary}

We have described the main ideas centered around Sharafutdinov's ray
transform which represents a generalization of the Radon transform to
tensor-field integrands. The tensorial character of the field to be
reconstructed manifests in a non-trivial kernel of the ray transform,
as a consequence of which the ray transform can be inverted only with
respect to the transverse components of the tensors, while the
longitudinal components are lost. We have provided an alternative
proof to the main theorem about the kernel of the ray transform as
given in Sharafutdinov's original work, which is simpler since the
scope of our proof is limited to those cases which are relevant for
the field of Integrated Photoelasticity. In order to illustrate the
merits of the ray transform as a tool for tensor reconstruction, a
simple example in Photoelasticity, namely the reconstruction of the
$\sigma_{33}$ component of the stress tensor inside a birefringent
medium from its horizontal ray transform, has been discussed. The
inversion of the ray transform for the transverse components of the
stress tensor together with the equilibrium conditions on
$\sigma_{ij}$ and the stress-free condition of the lateral boundary,
resulted in a well-posed Dirichlet problem for the component
$\sigma_{33}$ of the stress tensor which admits a unique
solution. Finally we suggested a non-linear inversion algorithm based
on the Newton-Kantarovich method with fixed derivative which might be
capable of utilizing the ray transform in the high-stress regime.

\begin{acknowledgements}
 The authors gratefully acknowledge the support of EPSRC grant
 \newline GR/86300/01.
\end{acknowledgements}

%
%





\end{article}
\end{document}